\documentclass[12pt]{article}

\usepackage{amssymb}
\usepackage{amsmath}
\usepackage{amsfonts}

\newcommand{\C}{\mathbb{C}}

\newcommand{\be}{\begin{equation}}
\newcommand{\bea}{\begin{eqnarray}}
\newcommand{\eea}{\end{eqnarray}}
\newcommand{\nn}{\nonumber}
\newcommand{\kt}{\rangle}
\newcommand{\br}{\langle}

\newcommand{\ed}{\end{document}}
\newcommand{\bbr}{\br\!\br}
\newcommand{\kkt}{\kt\!\kt}

\begin{document}

\title{Pseudo-Hermiticity versus $PT$ Symmetry III: Equivalence of pseudo-Hermiticity and the presence of antilinear symmetries}
\author{Ali Mostafazadeh\thanks{E-mail address: amostafazadeh@ku.edu.tr}\\ \\
Department of Mathematics, Ko\c{c} University,\\
Rumelifeneri Yolu, 80910 Sariyer, Istanbul, Turkey}
\date{ }
\maketitle

\begin{abstract}
We show that a diagonalizable (non-Hermitian) Hamiltonian $H$ is pseudo-Hermitian if and only if it has an antilinear symmetry, i.e., a symmetry generated by an antilinear operator. This implies that the eigenvalues of $H$ are real or come in complex conjugate pairs if and only if $H$ possesses such a symmetry. In particular, the reality of the spectrum of $H$ implies the presence of an antilinear symmetry. We further show that the spectrum of $H$ is real if and only if there is a positive-definite inner-product on the Hilbert space with respect to which $H$ is Hermitian or alternatively there is a pseudo-canonical transformation of the Hilbert space that maps $H$ into a Hermitian operator.
\end{abstract}

\baselineskip=24pt

\newpage
\section*{I.~Introduction}

The main reason for the recent interest in $PT$-symmetry \cite{pt} is that the eigenvalues of every $PT$-symmetric Hamiltonian are real or come in complex conjugate pairs. In particular, if the $PT$-symmetry is exact, the spectrum of the Hamiltonian is real. In Ref.~\cite{I}, we introduced the concept of a pseudo-Hermitian operator and showed that the remarkable spectral properties of the $PT$-symmetric Hamiltonians follow from their pseudo-Hermiticity. Under the assumption of the
diagonalizability (equivalently the existence of a complete biorthonormal set of energy eigenvectors),
we obtained in Ref.~\cite{II} a complete characterization of all the (non-Hermitian) Hamiltonian that have a real spectrum. Here we also pointed out that the spectral properties of the $PT$-symmetric Hamiltonians are common to all Hamiltonians  possessing an antilinear symmetry (a symmetry generated by an invertible antilinear operator.) Therefore, at least for the class of diagonalizable Hamiltonians presence of an antilinear symmetry implies pseudo-Hermiticity of the Hamiltonian. The primary purpose of the present article is to show that the converse of this statement holds as well. That is  pseudo-Hermiticity of a Hamiltonian implies the existence of an antilinear symmetry. A direct consequence of this result is that if the spectrum of the Hamiltonian is real, then the system has an
antilinear symmetry, $PT$-symmetry being the prime example.

The organization of the article is as follows. Section~II includes a brief review of the necessary results reported in the companion articles \cite{I,II}. Section~III examines anti-pseudo-Hermiticity (pseudo-Hermiticity with an antilinear automorphism.) Here we prove that every (non-Hermitian) diagonalizable Hamiltonian is anti-pseudo-Hermitian and that the pseudo-Hermiticity of the Hamiltonian implies the presence of an antilinear symmetry. Section~IV offers a description of the Hamiltonians with a real spectrum in terms of certain associated Hermitian operators. Section~V presents a summary of the main results and the concluding remarks.

Throughout this paper we shall consider (non-Hermitian) Hamiltonians $H$ that are diagonalizable
and have a discrete spectrum. As we explain below, this means that these Hamiltonians admit a complete biorthonormal set of eigenvectors $\{(|\psi_n,a\kt,|\phi_n,a\kt)\}$. The latter satisfy the following defining relations \cite{biortho}:
    \bea
    &&H |\psi_n,a\kt=E_n |\psi_n,a\kt ,~~~~H^\dagger|\phi_n,a\kt=E_n^*|\phi_n,a\kt,
    \label{bi-1}\\
    &&\br\phi_m,b|\psi_n,a\kt=\delta_{mn}\delta_{ab},
    \label{bi-2}\\
    &&\sum_n \sum_{a=1}^{d_n}|\psi_n,a\kt\br\phi_n,a|=1,
    \label{bi-3}
    \eea
where $n$ and $a$ are respectively the spectral and degeneracy labels, $d_n$ is the multiplicity
(degree of degeneracy) of $E_n$,  $\dagger$ and $*$ respectively denote the adjoint and complex-conjugate, $\delta_{mn}$ stands for the Kronecker delta function, and $1$ is the identity operator.
In view of Equations (\ref{bi-1}) -- (\ref{bi-3}), we also have
    \be
    H=\sum_n\sum_{a=1}^{d_n}E_n|\psi_n,a\kt\br\phi_n,a|,~~~~~~~~
    H^\dagger=\sum_n\sum_{a=1}^{d_n}E^*_n|\phi_n,a\kt\br\psi_n,a|.
    \label{HH}
    \end{equation}

In order to see the equivalence of the existence of a complete biorthonormal set of eigenvectors of $H$ and its diagonalizability, we note that by definition a diagonalizable Hamiltonian $H$ satisfies
$A^{-1}HA=H_0$ for an invertible linear operator $A$ and a diagonal linear operator $H_0$, i.e., there is an orthonormal basis $\{|n,\alpha\kt\}$ in the Hilbert space and complex numbers $E_n$ such that $H_0=\sum_n\sum_\alpha E_n |n,\alpha\kt\br n,\alpha|$. Then letting $|\psi_n,\alpha\kt:=A|n,\alpha\kt$ and $|\phi_n,\alpha\kt:=(A^{-1})^\dagger |n,\alpha\kt$, we can easily check that $\{|\psi_n,\alpha\kt,|\psi_n,\alpha\kt\}$ is a complete biorthonormal system for $H$. The converse is also true, for if such a system exists we may set $A:=\sum_n\sum_\alpha |\psi_n,\alpha\kt\br n,\alpha|$
for some orthonormal basis $\{|n,\alpha\kt\}$  and check that $A^{-1}=\sum_n\sum_\alpha
|n,\alpha\kt\br\phi_n,\alpha|$ and $A^{-1}HA=H_0$, i.e., $H$ is diagonalizable.

We would like to emphasize that the diagonalizability condition may be viewed as a physical requirement without which an energy eigenbasis would not exist. To our knowledge all known non-Hermitian Hamiltonians that are used in physical applications are diagonalizable and therefore admit a complete biorthonormal set of eigenvectors. This in particular includes all the Hermitian Hamiltonians as well as the non-Hermitian Hamiltonians used in ionization optics \cite{optics}, the study of dissipative systems and resonant states \cite{dissipation}, two-component formulation of the
minisuperspace quantum cosmology \cite{jmp98}, and also the $PT$-symmetric Hamiltonians whose spectral properties have been obtained using numerical methods.

\section*{II. Pseudo-Hermiticity}

Let $H:{\cal H}\to{\cal H}$ be a linear operator acting in a Hilbert space ${\cal H}$ and
$\eta:{\cal H}\to{\cal H}$ be a linear Hermitian automorphism (invertible transformation). Then
the $\eta$-pseudo-Hermitian adjoint of $H$ is defined by \cite{I}
    \be
    H^\sharp:=\eta^{-1}H^\dagger\eta.
    \label{sharp}
    \end{equation}
$H$ is said to be pseudo-Hermitian with respect to $\eta$ or simply $\eta$-pseudo-Hermitian if  $H^\sharp=H$. $H$ is said to be pseudo-Hermitian if it is pseudo-Hermitian with respect to some linear Hermitian automorphism $\eta$.

The basic properties of pseudo-Hermitian operators are discussed in Refs.~\cite{I,II}. Here we survey the properties that we shall make use of in this article. Let $H:{\cal H}\to{\cal H}$ be a diagonalizable linear operator. Then
    \begin{itemize}
    \item[--] $H$ is pseudo-Hermitian if and only if its eigenvalues are real or come in complex-conjugate pairs, \cite{I};
    \item[--] if $H$ is pseudo-Hermitian with respect to two linear Hermitian automorphisms
    $\eta_1$ and $\eta_2$, then $\eta_1^{-1}\eta_2$ generates a symmetry of $H$, i.e.,  $[H,\eta_1^{-1}\eta_2]=0$, \cite{I}.
    \end{itemize}

\section*{III. Anti-Pseudo-Hermiticity}

We first recall that a function $\tau:{\cal H}\to{\cal H}$ acting in a (complex) Hilbert space ${\cal H}$ is said to be an antilinear operator if for all $a,b\in\C$ and  $|\xi\kt,|\zeta\kt\in{\cal H}$,
    \be
    \tau(a|\xi\kt+b|\zeta\kt)=a^*\tau|\xi\kt+b^*\tau|\zeta\kt.
    \label{anti-1}
    \end{equation}
An antilinear operator $\tau:{\cal H}\to{\cal H}$ is said to be anti-Hermitian \cite{weinberg} if for all  $|\xi\kt,|\zeta\kt\in{\cal H}$,
    \be
    \br\zeta|\tau|\xi\kt=\br\xi|\tau|\zeta\kt.
    \label{anti-2}
    \end{equation}
    \begin{itemize}
    \item[] {\bf Definition~1:} A linear operator $H:{\cal H}\to{\cal H}$ acting in a Hilbert space     ${\cal H}$ is said to be anti-pseudo-Hermitian if there is an antilinear anti-Hermitian     automorphism $\tau:{\cal H}\to{\cal H}$ satisfying
    \be
    H^\dagger=\tau H \tau^{-1}.
    \label{p-h-anti}
    \end{equation}
    \end{itemize}

We begin our analysis by giving a characterization of antilinear anti-Hermitian operators with respect to which a given
linear operator is anti-pseudo-Hermitian.
    \begin{itemize}
    \item[] {\bf  Theorem~1:} Let ${\cal H}$ be a Hilbert space and $H:{\cal H}\to{\cal H}$ be a diagonalizable linear operator with a discrete spectrum and a complete biorthonormal set of eigenvectors $\{(|\psi_n,a\kt,|\phi_n,a\kt)\}$. Then $\tau:{\cal H}\to{\cal H}$ is an antilinear anti-Hermitian operator and $H$ is $\tau$-anti-pseudo-Hermitian if and only if there are symmetric invertible matrices $c^{(n)}$ with entries $c^{(n)}_{ab}$ such that for all $|\zeta\kt\in{\cal H}$,
    \be
    \tau|\zeta\kt=\sum_n\sum_{a,b=1}^{d_n}c^{(n)}_{ab}
    \br\zeta|\phi_n,a\kt\,|\phi_n,b\kt.
    \label{tau=c}
    \end{equation}
    \item[] {\bf Proof:} Suppose that $\tau:{\cal H}\to{\cal H}$ is a given antilinear anti-Hermitian operator and $H$ is $\tau$-anti-pseudo-Hermitian, i.e., (\ref{p-h-anti}) or equivalently
    \be
    H^\dagger\tau=\tau H
    \label{p-h-t}
    \end{equation}
holds. Letting both sides of (\ref{p-h-t}) act on $|\psi_n,a\kt$ and using (\ref{bi-1}) and (\ref{anti-1}),
we have
    \[H^\dagger(\tau|\psi_n,a\kt)=E_n^*(\tau|\psi_n,a\kt).\]
Comparing this equation with the second equation in (\ref{bi-1}), we find
    \be
    \tau|\psi_n,a\kt=\sum_{b=1}^{d_n} c^{(n)}_{ba}|\phi_n,b\kt,
    \label{x1}
    \end{equation}
where $c^{(n)}_{ab}$ are defined by
    \be
    c^{(n)}_{ab}:=\br\psi_{n,a}|\tau|\psi_n,a\kt.
    \label{c}
    \end{equation}
We can also express~(\ref{x1}) in the form
    \be
    \br\psi_m,b|\tau|\psi_n,a\kt=\delta_{mn}c^{(n)}_{ba}
    \label{x2}
    \end{equation}
Next note that because $\tau$ is an invertible operator, the matrix $c^{(n)}=(c^{(n)}_{ab})$ formed by
$c^{(n)}_{ab}$ is nonsingular. In fact, applying $\br\phi_{n,c}|\tau^{-1}$ to both sides of (\ref{x1})
and using (\ref{anti-1}) and the fact that $\tau^{-1}$ is also antilinear, we have
    \be
    ({c^{(n)}}^{-1})_{ab}=\br\phi_n,a|\tau^{-1}|\phi_n,b\kt^*.
    \label{c*}
    \end{equation}
Furthermore, in view of (\ref{c}) and (\ref{anti-2}), $c^{(n)}$ is a symmetric matrix. Now let $|\zeta\kt$ be an arbitrary element of
${\cal H}$ and use (\ref{x2}), (\ref{bi-3}) and (\ref{anti-2}) to compute
    \bea
    \sum_n\sum_{a,b=1}^{d_n}c^{(n)}_{ab} \br\zeta|\phi_n,a\kt\,|\phi_n,b\kt&=&
    \sum_{n,m}\sum_{a,b=1}^{d_n} |\phi_m,b\kt \br\zeta|\phi_n,a\kt \br\psi_n,a|\tau|\psi_m,b\kt \nn\\
    &=&\sum_m\sum_b |\phi_m,b\kt \br\zeta|\tau| \psi_m,b\kt \nn\\
    &=&\sum_m\sum_b |\phi_m,b\kt \br\psi_m,b|\tau|  \zeta\kt \nn\\
    &=&\tau|\zeta\kt.\nn
    \eea
This establishes (\ref{tau=c}). Next, suppose that $c^{(n)}$ are given invertible symmetric matrices and $\tau$ is defined by (\ref{tau=c}). Then the antilinearity of $\tau$ follows from the antilinearity of the inner-product in its first entry. The following simple calculation shows that $\tau$ is anti-Hermitian. For all $|\xi\kt,|\zeta\kt\in{\cal H}$,
    \bea
    \br\xi|\tau|\zeta\kt&=&\sum_n\sum_{a,b=1}^{d_n}c^{(n)}_{ab} \br\zeta|\phi_n,a\kt\br\xi|\phi_n,b\kt\nn\\
    &=&\sum_n\sum_{a,b=1}^{d_n}c^{(n)}_{ab} \br\zeta|\phi_n,b\kt \br\xi|\phi_n,a\kt\nn\\
    &=&\sum_n\sum_{a,b=1}^{d_n}c^{(n)}_{ab}\br\xi|\phi_n,a\kt \br\zeta|\phi_n,b\kt\nn\\
    &=&\br\zeta|\tau|\xi\kt,\nn
    \eea
where we used (\ref{tau=c}) and the fact that $c^{(n)}$ are symmetric. In order to establish the $\tau$-anti-pseudo-Hermiticity of $H$ we first observe that (\ref{tau=c}) implies
    \be
    \tau^{-1}|\zeta\kt=\sum_n\sum_{a,b=1}^{d_n} (c^{(n)})^{-1*}_{ab}\br\zeta|\psi_n,a\kt\,|\psi_n,b\kt.
    \label{t-inv}
    \end{equation}
This can be easily checked by applying $\tau$ to the right-hand side of (\ref{t-inv}) and using (\ref{tau=c}), (\ref{bi-2}), and (\ref{bi-3}) to show that the result is $|\zeta\kt$. Next, we note that applying both sides of (\ref{tau=c}) to $|\psi_n,a\kt$ we recover (\ref{x1}). Finally, we use  (\ref{tau=c}), (\ref{t-inv}), (\ref{anti-1}), (\ref{bi-1}), (\ref{HH}), and (\ref{x1}) to compute, for all $|\zeta\kt\in{\cal H}$,
    \bea
    \tau H \tau^{-1} |\zeta\kt&=&\tau \sum_n\sum_{a,b=1}^{d_n}
    {c^{(n)}}^{-1*}_{ab}\br\zeta|\psi_n,a\kt\,E_n |\psi_n,b\kt \nn\\
    &=& \sum_n\sum_{a,b=1}^{d_n}E_n^*
    {c^{(n)}}^{-1}_{ab}\br\zeta|\psi_n,a\kt^*\tau|\psi_n,b\kt \nn\\
    &=& \sum_n\sum_{a,b,c=1}^{d_n}E_n^*\br\psi_n,a|\zeta\kt
    c^{(n)}_{ca}{c^{(n)}}^{-1}_{ab}|\phi_n,c\kt \nn\\
    &=&\sum_n\sum_{b=1}^{d_n}E_n^*|\phi_n,b\kt\br\psi_n,b|\zeta\kt\nn\\
    &=&H^\dagger|\zeta\kt.\nn
    \eea
Therefore, $\tau H \tau^{-1}=H^\dagger$.~~$\square$
    \end{itemize}

We should emphasize that unlike the case of pseudo-Hermitian Hamiltonians, the anti-pseudo-Hermiticity does not restrict the energy spectrum. In fact, we can use Theorem~1 to prove the following.
    \begin{itemize}
    \item[] {\bf Corollary~1:} Every diagonalazable linear operator $H:{\cal H}\to{\cal H}$ with a discrete spectrum is anti-pseudo-Hermitian.
    \item[] {\bf Proof:} Let $\{(|\psi_n,a\kt,|\phi_n,a\kt)\}$ be a complete biorthonormal set of eigenvectors, and $\tau:{\cal H}\to{\cal H}$ be defined by (\ref{tau=c}) with $c^{(n)}=1$ for all $n$, i.e., for all $|\zeta\kt\in{\cal H}$,
    \be
    \tau|\zeta\kt:=\sum_n\sum_{a=1}^{d_n}
    \br\zeta|\phi_n,a\kt\,|\phi_n,a\kt.
    \label{tau}
    \end{equation}
Then according to Theorem~1, $\tau$ is an antilinear anti-Hermitian operator and $H$ is $\tau$-anti-pseudo-Hermitian.~~$\square$
    \item[] {\bf Corollary~2:} Every diagonalizable pseudo-Hermitian linear operator
$H:{\cal H}\to{\cal H}$ with a discrete spectrum has an antilinear symmetry.
    \item[] {\bf Proof:} Let $H$ be pseudo-Hermitian. Then according to Corollary~1 it is also anti-pseudo-Hermitian, i.e., there are a linear Hermitian automorphism $\eta:{\cal H}\to{\cal H}$ and an antilinear anti-Hermitian automorphism $\tau:{\cal H}\to{\cal H}$ such that
    \be
    \eta H\eta^{-1}=H^\dagger=\tau H\tau^{-1}.
    \label{HHH}
    \end{equation}
Hence, $[H,\eta^{-1}\tau]=0$. Clearly $\eta^{-1}\tau$ is an antilinear operator.~~$\square$
    \item[] {\bf Theorem~2:} Let $H:{\cal H}\to{\cal H}$ be a diagonalizable linear operator acting in a Hilbert space ${\cal H}$ with a discrete spectrum. Then the following are equivalent.
    \begin{enumerate}
    \item The eigenvalues of $H$ are real or come in complex-conjugate pairs.
    \item $H$ is pseudo-Hermitian.
    \item $H$ has an antilinear symmetry.
    \end{enumerate}
    \item[] {\bf Proof:} The equivalence of 1. and 2. was established in Ref.~\cite{I}; Corollary~2 shows that  2. implies 3.; the fact that 3. implies 1. follows from a simple calculation given in Ref.~\cite{II}.~~$\square$
    \end{itemize}

A class of $PT$-symmetric Hamiltonians are given by
    \be
    H=\frac{p^2}{2m}+V_1(x)+iV_2(x),
    \label{H=}
    \end{equation}
where $V_1$ and $V_2$ are respectively even and odd real-valued functions and the classical phase space is assumed to be real, i.e., $x$ and $p$ are the standard Hermitian operators representing the position and momentum of a particle of mass $m$. As we point out in \cite{I}, the Hamiltonian (\ref{H=}) is $P$-pseudo-Hermitian. It is also easy to check that it is $T$-anti-pseudo-Hermitian. The $P$-pseudo-Hermiticity and $T$-anti-pseudo-Hermiticity of this Hamiltonian implies its $P^{-1}T=PT$
symmetry. In general, there are $PT$-symmetric Hamiltonians $H$ that are neither $P$-pseudo-Hermitian nor $T$-anti-pseudo-Hermitian. According to Theorem~2, if we make the physical assumption that $H$ is diagonalizable, so that it admits a complete biorthonormal set of energy eigenvectors, then $H$ must be pseudo-Hermitian with respect to a linear Hermitian automorphism $\eta$. It turns out that the choice of $\eta$ is not unique.
But fixing an antilinear anti-Hermitian operator $\tau$ with respect to which $H$ is anti-pseudo-Hermitian (namely (\ref{tau=c})), we can express $\eta$ in terms of $PT$ and $\tau$ according to
    \be
    \eta=\tau PT.
    \label{eta=}
    \end{equation}
One can easily check that $PT$ symmetry ($[PT,H]=0$) and anti-pseudo-Hermiticity (\ref{p-h-anti})
imply pseudo-Hermiticity of $H$ with respect to (\ref{eta=}).

Next we consider a general diagonalizable Hamiltonian $H$ with a discrete spectrum and a symmetry generated by a general antilinear operator $X$,
    \be
    [H,X]=0.
    \label{sym}
    \end{equation}
The antilinearity of $X$ implies $\eta$-pseudo-Hermiticity of $H$ with respect to some linear Hermitian automorphism $\eta$. The anti-pseudo-Hermiticity of $H$ with respect to an antilinear automorphism of the form~(\ref{tau=c}) always holds. Hence Eqs.~(\ref{HHH}) are valid. Taking the adjoint of both sides of (\ref{sym}) and making use of (\ref{HHH}), we can easily show that
$X^\sharp_\eta:=\eta^{-1}X\eta$ and $X^\sharp_\tau:=\tau^{-1}X\tau$ commute with $H$, i.e.,
they generate antilinear symmetries of the system as well.

\section*{IV. Non-Hermitian Hamiltonians with a Real Spectrum}

We first recall the following results which we reported in \cite{I,II}.
    \begin{itemize}
    \item[1.] The (indefinite) inner-product defined by
    \be
    \forall |\xi\kt,|\zeta\kt\in {\cal H}, ~~~~~~~
    \bbr\xi|\zeta\kkt:=\br\xi|\eta|\zeta\kt,
    \label{inner}
    \end{equation}
    is invariant under the evolution generated by an $\eta$-pseudo-Hermitian Hamiltonian $H$,   \cite{I}. It is also easy to check that such a Hamiltonian is Hermitian with respect to the     (indefinite) inner-product~(\ref{inner}). See also \cite{japaridze}.
    \item[2.] A diagonalizable (non-Hermitian) Hamiltonian  has a real spectrum if and only if it is    pseudo-Hermitian with respect to a linear Hermitian automorphism of the form
        \be
        \eta=OO^\dagger,
        \label{real}
        \end{equation}
    where $O:{\cal H}\to{\cal H}$ is a linear automorphism, \cite{II}.
    \end{itemize}
These statements suggest the following characterization of the (non-Hermitian) Hamiltonians with a real spectrum. See also \cite{kretschmer}.
    \begin{itemize}
    \item[] {\bf Theorem~3:} A diagonalizable Hamiltonian $H$ acting in a Hilbert space
    ${\cal H}$ has a real spectrum if and only if there is a positive-definite inner product on
    ${\cal H}$ with respect to which $H$ is Hermitian.
    \item[] {\bf Proof:} Suppose $H$ has a real spectrum so that it is
    $OO^\dagger$-pseudo-Hermitian for a linear automorphism $O:{\cal H}\to{\cal H}$. Then the
    inner-product (\ref{inner}) with $\eta=OO^\dagger$ is clearly a positive-definite inner-product
    with respect to which $H$ is Hermitian. Conversely, suppose that there is a positive-definite
    inner-product $(~,~)$ with respect to which $H$ is Hermitian. Then treating the spectral problem
    for $H$ in the Hilbert space ${\cal H}$ with the inner-product $(~,~)$, we find that $H$ has a
    real spectrum.
    \item[] {\bf Corollary:} Suppose that $H$ has an antilinear symmetry $X$. If $X$
    is an exact symmetry of $H$, then there is a positive-definite inner product on ${\cal H}$ with     respect to which $H$ is Hermitian.~~$\square$
    \item[] {\bf Proof.} Exactness of an antilinear symmetry implies reality of the spectrum of $H$,
    \cite{II}. The conclusion then follows from Theorem~3.~~$\square$
    \end{itemize}

Next we give an alternative and in a sense equivalent characterization of the (non-Hermitian) Hamiltonians with a real spectrum.
    \begin{itemize}
    \item[] {\bf Definition~2:} Consider a quantum system with the Hilbert space ${\cal H}$ and the     Hamiltonian $H:{\cal H}\to{\cal H}$. Then a linear automorphism $A:{\cal H}\to{\cal H}$ is  said to be a {\em pseudo-canonical transformation} for the system if for all $|\zeta\kt\in{\cal     H}$ the transformation
    \be
    |\zeta\kt\to|\tilde\zeta\kt:=A|\zeta\kt
    \label{p-c}
    \end{equation}
    leaves the Schr\"odinger equation,
    \be
    i\frac{d}{dt}|\psi(t)\kt=H|\psi(t)\kt,
    \label{sch-eq}
    \end{equation}
    form-invariant. A unitary pseudo-canonical transformation is called a (quantum) canonical   transformation \cite{nova}.
    \end{itemize}
Clearly the defining condition for a pseudo-canonical transformation implies the following transformation rule for the Hamiltonian.
    \be
    H\to\tilde H:=AHA^{-1}+i\dot A A^{-1},
    \label{trans}
    \end{equation}
where a dot denotes a time-derivative. For a time-independent pseudo-canonical transformation $A$, the second term on the right-hand side of (\ref{trans}) drops and $H$ transforms as
    \be
    H\to\tilde H:=AHA^{-1}.
    \label{trans-0}
    \end{equation}
    \begin{itemize}
    \item[] {\bf Theorem~4:} A diagonalizable Hamiltonian $H$ has a real spectrum if and only if
    there is a pseudo-canonical transformation that maps $H$ into a     Hermitian operator.
    \item[] {\bf Proof:} Suppose that $H$ has a real spectrum. Then it is
    $OO^\dagger$-pseudo-Hermitian for a linear automorphism $O:{\cal H}\to{\cal H}$, i.e.,  $H^\dagger=OO^\dagger H(OO^\dagger)^{-1}$. Let $A:=O^\dagger$, then in view of  (\ref{trans-0}) and the preceding equation, we have
        \[ \tilde H^\dagger=(AHA^{-1})^\dagger=
        (A^{-1})^\dagger H^\dagger A^\dagger=
        (A^{-1})^\dagger A^\dagger A H A^{-1} (A^\dagger)^{-1}A^\dagger
        =AHA^{-1}=\tilde H.\]
    Hence the transformed Hamiltonian is Hermitian. Conversely suppose that there is a
    pseudo-canonical transformation $A:{\cal H}\to{\cal H}$ under which $H$ transforms
    to a Hermitian Hamiltonian $\tilde H$ and let $O:=A^\dagger$. Then using (\ref{trans-0}) and
    $\tilde H^\dagger=\tilde H$, we have
        \[ OO^\dagger H(OO^\dagger)^{-1}=A^\dagger A A^{-1}\tilde H A (A^\dagger A)^{-1}
            = A^\dagger \tilde H^\dagger (A^\dagger)^{-1}=(A^{-1}\tilde H A)^\dagger=
            H^\dagger.\]
    Therefore, $H$ is $OO^\dagger$-pseudo-Hermitian, and its spectrum is real.~~$\square$
    \item[] {\bf Corollary:} Suppose that $H$ has an antilinear symmetry $X$. If $X$
    is an exact symmetry of $H$, then there is a pseudo-canonical transformation that maps $H$ into     a Hermitian operator.
    \item[] {\bf Proof.} Exactness of an antilinear symmetry implies reality of the spectrum of $H$,
    \cite{II}. The conclusion then follows from Theorem~4.~~$\square$
    \end{itemize}

\section*{V. Discussion and Conclusion}
In this article we established the equivalence of the notion of pseudo-Hermiticity and presence of an antilinear symmetry for the class of diagonalizable  (non-Hermitian) Hamiltonians. This required the study of pseudo-Hermiticity with respect to antilinear anti-Hermitian automorphisms. It turned that the latter does not restrict the choice of the Hamiltonian and such antilinear automorphisms always exist. In fact, we obtained the general form of these automorphisms. For a fixed complete biorthonormal eigenbasis, they are determined in terms of a sequence of complex symmetric matrices $c^{(n)}$. The choice of unity for all these matrices leads to a canonical antilinear anti-Hermitian automorphism, namely (\ref{tau}). Under an invertible  transformation $u$ of the basis,
    \[ |\psi_n,a\kt\to \sum_{b=1}^{d_n} u_{ba}|\psi_n,b\kt,~~~~~
        |\phi_n,a\kt\to\sum_{b=1}^{d_n} (u^{-1\dagger})_{ba}|\phi_n,b\kt,\]
that preserves its completeness and biorthonormality, the matrices $c^{(n)}$ transform according to
    \[ c^{(n)}\to u^{*t}c^{(n)}u^{*}=u^{\dagger}c^{(n)}u^{\dagger t},\]
where $^t$ denotes the transpose. We can transform to a basis where a general $\tau$ has the canonical form (\ref{tau}) if we can find invertible matrices $v=u^{-1\dagger}$ satisfying $c^{(n)}=vv^t$. As
shown in \cite{factor} this is always possible. Therefore, up to the choice of the biorthonormal
eigenbasis, $\tau$ is actually unique.

A simple consequence of our findings is that the reality of the spectrum of a Hamiltonian implies the presence of an antilinear symmetry. In view of the proof of Corollary~2 and Eq.~(\ref{tau=c}) of this
article and Eq.~(23) of \cite{I}, we have in fact an explicit expression for the generator of such a symmetry in terms of the biorthonormal eigenvectors of the Hamiltonian. We also gave two characterizations of Hamiltonians with real spectrum. These characterizations show how a Hamiltonian
with a real spectrum may be related to an associated Hermitian Hamiltonian. Another simple implication of our analysis is that every Hermitian Hamiltonian has an antilinear symmetry.

\section*{Acknowledgment}
This project was supported by the Young Researcher Award Program (GEBIP) of the Turkish Academy of Sciences.

\ed